\documentclass[a4paper,11pt]{report}
\usepackage[T1]{fontenc}
\usepackage[utf8]{inputenc}
\usepackage{lmodern}
\usepackage{graphicx}
\usepackage{amsmath}
\usepackage{amssymb}
\usepackage{amsfonts}
\usepackage{latexsym}
\usepackage{fancyhdr}
\usepackage{graphicx}
\usepackage{color}
\usepackage{graphics}
\usepackage{color}

\begin{document}

\begin{center}
{\bf Non-relativistic Reduction of Spinors, New Currents and their Algebra} \\ 
\vspace{3 mm}
Rabin Banerjee{\footnote{rabin@bose.res.in}} \\
S. N. Bose National Centre for Basic Sciences \\
Block - JD, Sector – III, Salt Lake City, Kolkata - 700 106, India \\
and \\
Debashis Chatterjee {\footnote{debachatter@gmail.com}}\\
Vijaygarh Jyotish Ray College, 8/2 Vijaygarh, Kolkata - 700 032, India \\
\end{center}
\vspace{3 mm}

{\bf Abstract :} A specific mapping is introduced to reduce the Dirac action to the non-relativistic (Pauli - Schr\"odinger) action for spinors. 
Using this mapping, the structures of the vector and axial vector currents in the non-relativistic theory are obtained. The implications of the relativistic 
Ward identities in the non-relativistic limit are discussed. A new non-abelian type of current in the Pauli - Schr\"odinger theory is obtained. As we show, this is essential for the closure of the algebra among the usual currents. The role of parity in the non-relativistic theory is also discussed.
\newpage
\begin{center}
\bf {Section I. Introduction}
\end{center}

The low energy effective description of a system usually requires the study of a non-relativistic (NR) field theory. In general, however such theories and their symmetries are difficult to handle due to the occurrence of a universal time so that a systematic covariant formulation is no longer available. Possible ways are to construct the galileo invariant wave equation adopting Dirac's proceedure \cite{LL} or, alternatively,  to abstract the NR limit of the corresponding relativistic theory by developing a systematic algorithm. This approach has found applications in various contexts; for example, in fluid dynamics \cite{Horv, Kami}, in identifying possible dark matter candidates in particle physics models \cite{Braa, Namj}, understanding NR diffeomorphism \cite{Jens, Bane1, Bane2} 
and other assorted phenomena \cite{Duva, Bane3, Ichi}. 

At the kinematic level there is the group contraction method that takes a group associated with a relativistic theory to the group corresponding to the NR theory in a certain limit.  A typical example is the contraction of the Poincare group to the Galilean group. Such contraction techniques, however, fail to provide insights or connections into the dynamical aspects of the concerned theories. Thus if we want to obtain the Schr\"odinger theory as a NR limit of relativistic scalar field theory, the group contraction method is no longer helpful. This issue has been discussed at some length in \cite{Bane3}.  For example, the Schr\"odinger theory cannot be obtained from the real Klein-Gordon theory simply because the $U(1)$ invariance of the former is not there  in the latter. One may then try a complex scalar  theory but other complications arise and the agreement of symmetries becomes problematic.  This is basically tied to the fact that the complex scalar theory, unlike the Schr\"odinger theory, is a second order theory.  This prompts another approach, the so called Eisenhart lift approach \cite{Eise, Duval}.  In this case the scalar theory is expressed in light cone coordinates which essentially converts it into a first order system. Then the passage to the Schr\"odinger theory may be achieved \cite{Bane3}.
There is a reasonable amount of literature \cite{Bane2, Namj, Bane3} in discussing the transition of the scalar theory to the Schr\"odinger theory, both in flat and curved backgrounds.

As is known, Pauli modified the Schr\"odinger theory to introduce spin in a NR context which is called the Pauli-Schr\"odinger (PS) model. It is therefore expected that this model should be recovered from the standard Dirac theory. In this specific case both the Dirac and PS  theories are first order, so the additional complexities arising in the scalar case are avoided. Naturally it is also not necessary to pass to the light cone formulation. The transition from the Dirac theory to the PS theory has also been treated in the literature, albeit with the motivation of consistency checks \cite{Gurt, Nowa, Shik}. Also the dimensional descent approach has been discussed \cite{Duval}. Non-relativistic limit in the generic background has also been considered \cite{Fuin}.
 
In this paper we will discuss the NR reduction of the Dirac theory to the PS theory in full details, highlighting many aspects that have not been considered previously. In section II, a specific mapping is introduced to achieve this reduction. Explicitly, we show the mapping among the field equations which is extended to include the respective actions.   In section III, an analysis of the Noether currents and their conservation has been performed. Both vector and axial vector currents are treated.  The axial case, in particular, shows how the role of $\gamma_5$ is translated in the PS theory, despite the fact that it cannot be introduced in a conventional way in the NR case. The axial Ward identity for the Dirac theory actually translates to an identity in the PS theory. This is distinct from the standard way of obtaining the Ward identities by using the equations of motion. A detailed account of the current algebra is presented in section IV. We now have a surprise. The algebra does not close hinting at the presence of some new currents. We are able to identify the source of these new currents. The point is that in contrast to the pure Schr\"odinger theory involving a single component object, the PS theory admits an extra non-abelian type symmetry transformation leading to these new currents. The complete algebra involving the standard currents and the new currents now closes. In section V we have discussed the role of parity transformations in the Dirac and PS theories. These transformations are consistently related using our mapping. Section VI contains our conclusions.

\begin{center}
\bf {Section II. Field Equation, Lagrangian and the Space-time Generators}
\end{center}

\noindent {\bf IIA. The Free Field Equations}  
\bigskip

Here we briefly review the nonrelativistic reduction of the Dirac equation. We start with the standard form of the Dirac equation :
\begin{equation}\label{Dirac}
i\hbar {\frac{\partial \psi} {\partial t}} = (- i\hbar c \vec \alpha \cdot \vec \nabla + \beta m c^2) \psi, 
\end{equation}
where $\alpha$ and $\beta$ are the well known Dirac matrices: \begin{equation}\label{alpha}
\vec \alpha = \begin{pmatrix}
0 &  {\vec \sigma} \\
{\vec \sigma} & 0
\end{pmatrix} \qquad
\beta = \begin{pmatrix}
I &0  \\
0 & -I
\end{pmatrix}
\end{equation}
$\sigma_i$'s being the three Pauli matrices.

 If we factor out the oscillation due to the rest mass by substituting :
\begin{equation} \label{psi0}
\psi = e^{- im c^2 t/ \hbar} \psi_0, 
\end{equation}
then the time evolution of $\psi_0$ is given by the equation :  
\begin{equation}\label{psi0eq}
i\hbar \bigg\{ {\partial \psi_0 \over \partial t} - (im c^2/ \hbar) \psi_0 \bigg \} = -  i\hbar c \alpha_i  {\partial \psi_0 \over \partial x_i} + m c^2 (\beta \psi_0). 
\end{equation}
Next, we split the four component spinor $\psi_0$ into two two-component spinors $\xi_0$ and $\eta_0$ :
\begin{equation}\label{xi0eta0}
\psi_0 = \begin{pmatrix}
\xi_0 \\
\eta_0
\end{pmatrix},
\end{equation}
so that (\ref{psi0eq}) reads :
\begin{equation}\label{psi0eq1}
i\hbar  {\partial \over \partial t} 
\begin{pmatrix} \xi_0 \\ \eta_0 \end{pmatrix} 
+ m c^2 \begin{pmatrix} \xi_0 \\ \eta_0 \end{pmatrix} = 
- i \hbar c \begin{pmatrix} 0 & \sigma^i \\
\sigma^i& 0 \end{pmatrix}
\begin{pmatrix} \partial_i \xi_0 \\
\partial_i \eta_0 \end{pmatrix} 
+ mc^2 \begin{pmatrix} \xi_0 \\ - \eta_0 \end{pmatrix}. 
\end{equation}
which splits into the pair of equations :
\begin{equation}\label{xi0eq}
i\hbar {\partial \xi_0 \over \partial t} = -  i\hbar c \sigma_i  {\partial \eta_0 \over \partial x_i}, 
\end{equation}
\begin{equation}\label{eta0eq}
i\hbar {\partial \eta_0 \over \partial t} = -  i\hbar c \sigma_i  {\partial \xi_0 \over \partial x_i} - 2m c^2 \eta_0, 
\end{equation}
In the large `$c$' limit,  (\ref{eta0eq}) reduces to,
\begin{equation}\label{eta-xi}
\eta_0 = - { i\hbar \over 2 m c} \sigma_i  \partial_i \xi_0 .
\end{equation}

Substituting the expression for $\eta_0$  from(\ref{eta-xi}) back in (\ref{xi0eq}), we obtain :
\begin{equation}\label{PSxi0}
i\hbar {\partial \xi_0 \over \partial t} = -{\hbar^2 \over 2 m} (\sigma_i \partial_i) (\sigma_k \partial_k) \xi_0. 
\end{equation}
Eq.(\ref{PSxi0}) was originally proposed by Pauli as an extension of Schr\"odinger equation to
incorporate spin. It was earlier obtained in \cite{LL},  mimicking the original Dirac construction. Also, the same result follows by extending the method of \cite{Duva}, beginning with the massless Dirac equation in one dimension higher and making a light like reduction \cite{Duval}. Using :
\begin{equation} \label{sigmalg}
\sigma_i\sigma_k={1 \over 2}\{ \sigma_i, \sigma_k \} + {1 \over 2} [\sigma_i, \sigma_k]
= \delta_{ik} + i\epsilon_{ikl}\sigma_l,
\end{equation}
the above equation reduces to the Schr\"odinger type equation :
\begin{equation}\label{Schxi0}
i\hbar {\partial \xi_0 \over \partial t} = - {\hbar^2 \over 2 m} \nabla^2 \xi_0,
\end{equation}
where $\xi_0$ is now a two-component spinor. 

Although (\ref{PSxi0}) and (\ref{Schxi0}) are algebraically equivalent, the role of spin is manifest only in (\ref{PSxi0}). 
One might wish to go one step further. Keeping the next term in large `$c$' limit, we find, 
\begin{equation}\label{Schxi01}
i\hbar {\partial \xi_0 \over \partial t} = - {\hbar^2 \over 2 m} \nabla^2 \xi_0 - 
{\hbar^4 \over 8 m^3 c^2}\nabla^4 \xi_0. 
\end{equation}
The second term on the right hand side of (\ref {Schxi01}) is the familiar relativistic correction to the 
Pauli-Schrödinger equation, that finds application, for example, in the hydrogen atom problem. The above equation is a differential representation of the non-relativistic energy operator, subtracting out the rest energy, 
\begin{equation}
\sqrt{c^2p^2 + m^2 c^4} - mc^2 = {p^2 \over 2m} - {p^4 \over 8m^3c^2} + 
 {\cal O}({1 \over c^4}). 
\end{equation}

\bigskip

\bigskip
\noindent {\bf IIB. The Negative Energy Solutions} 
\bigskip

Dirac equation, as we know permits negative energy solutions ($ E = - \sqrt{c^2p^2 + m^2 c^4}$), the energy spectrum being bounded {\it above} by $E_{max} = -m c^2$. To obtain the non-relativistic limit appropriate for these solutions, we replace (\ref{psi0}) by the substitution :
\begin{equation} \label{psi0-}
\psi = e^{+ im c^2 t/ \hbar} \psi_0. 
\end{equation}
The roles of $\xi_0$ and $\eta_0$ are now reversed. Following identical steps as before,  we obtain, 
\begin{equation}\label{xi-eta}
\xi_0 = { i\hbar \over 2 m c} \sigma_i  \partial_i \eta_0 .
\end{equation}
and,
\begin{equation}\label{Scheta0}
i\hbar {\partial \eta_0 \over \partial t} = + {\hbar^2 \over 2 m} \nabla^2 \eta_0, 
\end{equation}
which is Schr\"odinger equation with a negative mass. Let us keep in mind, that in the non-relativistic limit : $p \rightarrow 0$ for the negative energy solutions, $E \rightarrow - m c^2$, which can be interpreted as the rest energy of a particle having a negative rest mass.  

\bigskip
\noindent {\bf IIC. The Lagrangian} 
\bigskip

Despite an agreement among the equations of motions, there may be a mismatch between the corresponding actions. This, however, does not happen here. Let us start from the usual Dirac theory :
\begin{eqnarray}
{\cal L}^D & = & \bar \psi \bigg [i\hbar c\gamma^0 \partial_0 \psi + i\hbar c\gamma^i
\partial_i \psi  - m c^2 \psi \bigg ] \label{LDgamma}\\
& = & \psi^ \dagger\bigg [i\hbar {\partial \psi \over \partial t} + i\hbar c\alpha_i
{\partial \psi \over \partial x_i}  - m c^2 (\beta\psi)\bigg ],\label{LDalpha}
\end{eqnarray} 
where the various constants are chosen so that the Lagrangian density has the dimension of energy density. Expressed in terms of the 2-component spinors, ${\cal L}^D$ assumes the form : 
\begin{equation}\label{LDxi}
{\cal L}^D = i\hbar \bigg\{ \xi^ \dagger {\partial \xi \over \partial t} + \eta^ \dagger {\partial \eta \over \partial t} \bigg \} 
+ i\hbar c \bigg\{ \xi^ \dagger \sigma_i {\partial \eta \over \partial x_i}  + \eta^ \dagger \sigma_i {\partial \xi\over \partial x_i} \bigg \} - m c^2 \{ \xi^\dagger \xi - \eta^\dagger \eta \}.
\end{equation}
Making the substitution (\ref{psi0}), which may be equivalently written as :
\begin{equation}\label{xi0}
\xi = e^{- im c^2 t/ \hbar} \xi_0 , \quad \eta = e^{- im c^2 t/ \hbar} \eta_0 ,
 \end{equation}
the following expression is obtained : 
\begin{equation}\label{LDxi0}
{\cal L}^D = i\hbar \bigg\{ \xi_0^ \dagger {\partial \xi_0 \over \partial t} + \eta_0^ \dagger 
{\partial \eta_0 \over \partial t}  \bigg \} 
+ i\hbar c \bigg\{ \xi_0^ \dagger \sigma_i {\partial \eta_0 \over \partial x_i}  + \eta_0^ \dagger \sigma_i {\partial \xi_0\over \partial x_i} \bigg \} + (2m c^2) \eta_0^\dagger \eta_0 
\end{equation}
If we now replace the field $\eta_0$ by its expression (\ref{eta-xi}) as before and work out the sigma algebra, the Lagrangian density (\ref{LDxi0}), in the leading order approximation $c \rightarrow \infty $, leads to the Pauli-Schr\"odinger theory :
\begin{equation}\label{LPSxi0}
{\cal L}^{PS} = \bigg [i\hbar \xi_0^\dagger {\partial \xi_0 \over  \partial t}+ {\hbar^2 \over 2m}\xi_0^\dagger \nabla^2 \xi_0 \bigg ]. 
\end{equation}

One may similarly verify that such an equivalence holds at the hamiltonian level also, i.e. the various space time generators for the Pauli-Schroedinger theory may be reproduced from the Dirac theory in the appropriate limit.

\begin{center}
{\bf Section III. Noether Currents, their Conservation and Current Algebra}
\end{center}

In this section, we discuss the structures of the vector current, axial current and their divergences in the non-relativistic theory, starting from their relativistic counterparts. While the results for the vector case are reproduced by directly starting from the Pauli-Schr\"odinger Lagrangian, the same is {\it not} true for the axial case, due to the absence of $\gamma_5$ in the non-relativistic theory.

\bigskip
\noindent {\bf IIIA. The Vector Current and its Conservation} 
\bigskip

The Dirac Lagrangian, as we know, is symmetric under the ‘global’ gauge transformation : $\psi \rightarrow e^{-i\alpha } \psi$ and the corresponding Noether current density, defined as :
\begin{equation}\label{JDpsi}
J^{\mu D} = {1 \over (\hbar \alpha)}\>{\partial {\cal L} \over \partial (\partial_\mu \psi)} \delta \psi = 
c \bar \psi \gamma^\mu \psi
\end{equation}
is conserved. In three-vector notation :
\begin{equation}\label{JD0}
J_0^D = c\psi ^\dagger \psi, \quad  j_i^D = c \psi^\dagger \alpha_i \psi.
\end{equation}
The Pauli-Schr\"odinger theory (\ref{LPSxi0}) also respects the same gauge symmetry mentioned above and the corresponding charge and current density are obtained as :
\begin{eqnarray}
J_0^{PS} & = &{1 \over (\hbar \alpha)}\> {\partial {\cal L}_{PS} \over \partial(\partial_0 \xi_0)}\delta \xi_0 = c \xi_0 ^\dagger \xi_0, \label{J0PSpsi}\\  
J_i^{PS} & = &{1 \over (\hbar \alpha)}\> {\partial {\cal L}_{PS} \over \partial(\partial_i \xi_0)}\delta \xi_0 = {i\hbar \over 2m_0} \{ (\partial_i \xi_0 ^\dagger) \xi_0 -  \xi_0 ^\dagger (\partial_i \xi_0) \}.\label{JiPSpsi}
\end{eqnarray}
We propose to investigate, whether $J_0^D$ and $J_i^D$ (\ref{JD0}), obtained from the Dirac Lagrangian, reduce to their counterparts $J_0^{PS}$ and 
$J_i^{ PS}$ (\ref{J0PSpsi}) and (\ref{JiPSpsi}), in the non-relativistic limit. 
Making the substitution (\ref{psi0}) and re-writing in terms of the two-component spinors as usual, $J^0_D$ becomes :
\begin{equation} \label{JD0xi0}
J_0^D = c (\xi_0 ^\dagger \xi_0 + \eta_0 ^\dagger \eta_0).
\end{equation}
The second term, as indicated by (\ref{eta-xi}), is 
suppressed compared to the first, so that,
\begin{equation} \label{JD0xi01}
J_0 = \lim_{c \rightarrow \infty} J_0^D = c \xi_0 ^\dagger \xi_0.
\end{equation}
We thus find a complete agreement of $J_0$ with $J_0^{PS}$ in form. Similar 
operations reduce $J_i^D$ to 
\begin{equation} \label{JDixi0}
J_i = \lim_{c \rightarrow \infty} J_i^D = c (\xi_0 ^\dagger \sigma_i \eta_0 + \eta_0 ^\dagger\sigma_i \xi_0).
\end{equation}
As we substitute $\eta_0$ from (\ref{eta-xi}), we find :
\begin{equation}\label{JDixi01}
J_i = {i\hbar \over 2m} \bigg\{ {\partial \xi_0 ^\dagger \over \partial x_k} \sigma_k \sigma_i \xi_0 -  \xi_0 ^\dagger \sigma_i \sigma_k {\partial \xi_0 \over \partial x_k} \bigg \}.
\end{equation}
Making use of the sigma algebra (\ref{sigmalg}), we find that $J^i$ breaks up into two pieces :
 \begin{eqnarray} 
J_i& = & {i\hbar \over 2m} \bigg\{ {\partial \xi_0 ^\dagger \over \partial x_i} \xi_0 -  \xi_0 ^\dagger{\partial \xi_0 \over \partial x_i} \bigg \} +
{\hbar \over 2m}\bigg\{ \epsilon_{ikl}{\partial \over \partial x_k} (\xi_0 ^\dagger \sigma_l \xi_0) \bigg \} \\ \label{Ji1+2} 
& = &J_i^{PS} +
{\hbar \over 2m}\bigg\{ \epsilon_{ikl}{\partial \over \partial x_k} (\xi_0 ^\dagger \sigma_l \xi_0) \bigg \}.
\end{eqnarray}
Thus the non-relativistic limit of $J_i^D $ differs from the corresponding Pauli-Schr\"odinger current by a spin dependent term. If we extend our free theory to a case where a magnetic field is present, then the Dirac Lagrangian density \ref{LDgamma}), gets modified to : 
 \begin{equation}\label{LDA}
{\cal L}^D = \bar \psi \Big[i\hbar \gamma^0 {\partial \psi \over \partial t} + c\gamma^i
 (i\hbar {\partial \psi \over \partial x^i}+eA_i \psi) - m c^2 \psi\Big]
\end{equation}
The interaction term : $(ec \> \bar \psi \gamma^i \psi) A_i$ generates an extra piece in the action : 
\begin{eqnarray}
{e\hbar \over 2m}\bigg\{ \epsilon_{ikl}{\partial \over \partial x_k} (\xi_0 ^\dagger \sigma_l \xi_0) \bigg \}A_i \> d^3x 
& = &-{e\hbar \over 2m} \int \epsilon_{ikl}{\partial A_i \over \partial x_k} 
(\xi_0 ^\dagger \sigma_l \xi_0) \> d^3x \nonumber\\  
& = &-{e\hbar \over 2m}\int (\xi_0 ^\dagger \sigma_l \xi_0) B_l \> d^3x.\label{sigma.B}
\end{eqnarray} 
The last expression represents the well-known coupling of the spin magnetic moment of the fermion with the magnetic field. This is the origin of the additional term in the current (\ref{Ji1+2})  

As far as the conservation of the vector current (in the non-relativistic limit) is concerned, the second term in the right hand side of (\ref{Ji1+2}) does not contribute in $\partial_i J_i$, since it is the curl of a vector. Hence $\partial_i J_i$ equals $\partial_i J_i^{PS}$. Similarly, $J_0$ assumes a form identical to $J_0^{PS}$, as shown in (\ref{JD0xi01}). Therefore, the conservation of this current can be demonstrated in the usual way, with the help of the Pauli-Schr\"odinger equation. 

  It is useful to mention that the extra term in the vector current $J_i$ generated by the nonrelativistic reduction process may also be obtained directly from the Pauli-Schr\"odinger theory by reexpressing the Lagrangian obtained from (\ref{LPSxi0})  as :
\begin{equation}\label{LPSnew}
 L^{PS} = \int \bigg [i\hbar \xi_0^\dagger {\partial \xi_0 \over  \partial t}+ {\hbar^2 \over 2m}\xi_0^\dagger (\sigma_i \partial_i)(\sigma_k \partial_k)\xi_0 \bigg ] \> d^3x. 
\end{equation}
 The expression can be converted through an
integration by parts, to the form :
\begin{equation}\label{LPS1new}
L_1^{PS} = \int \bigg [i\hbar \xi_0^\dagger {\partial \xi_0 \over  \partial t}- {\hbar^2 \over 2m}(\partial_i \xi_0^\dagger)\sigma_i \sigma_k (\partial_k\xi_0) \bigg ] \> d^3x. 
\end{equation} 
This Lagrangian is also symmetric under the global transformation $ \xi_0 \rightarrow e^{-i\alpha} \xi_0 $. The Noether current corresponding to this symmetry matches exactly with (\ref{JD0xi01}) and (\ref{Ji1+2}).
   
\bigskip
\noindent {\bf  IIIB. The Axial Current and its Divergence}
\bigskip

The kinetic part of the Dirac Lagrangian respects another symmetry, which however is broken by the mass term. This is the chiral symmetry : $\psi \rightarrow e^{-i\alpha \gamma_5} \psi$. The corresponding Noether current has a non-zero divergence, proportional to the fermion mass.
\begin{equation} \label{PCAC}
\partial_\mu J^{\mu D}_5 = \partial_\mu \{ c\> \bar \psi \gamma^\mu\gamma_5 \psi \} = {2im c^2 \over \hbar}\bar \psi \gamma_5 \psi.
\end{equation}
The Schr\"odinger or the Pauli-Schr\"odinger theory, on the other hand, has no chiral symmetry, simply because the chirality operator $\gamma_5$ is not defined here. However, we can take the non-relativistic limit of the chiral current of the Dirac theory and thus find an analogue of the chiral current in the Pauli-Schr\"odinger theory.
Introducing the fields $\xi_0$ and $\eta_0$ as before, $J^0_{5D}$ and $J^i_{5D}$
assume the forms :
\begin{eqnarray}
J_{05}^D & = & c (\xi_0^\dagger \eta_0 + \eta_0^\dagger \xi_0) \label{J05}\\
J_{i5}^D & = & c (\xi_0^\dagger \sigma_i \xi_0 + \eta_0^\dagger \sigma_i \eta_0).\label{Ji5}
\end{eqnarray}
We also re-write the right hand side of (\ref{PCAC}) (let us call it $J_{5D}$) as : 
\begin{equation}\label{J5}
J_5^D = {2im c^2 \over \hbar}\bar \psi \gamma_5 \psi = {2im c^2 \over \hbar} (\xi_0^\dagger \eta_0 - \eta_0^\dagger \xi_0)
\end{equation}
Now we observe that the second term in the expression for $J_{i5}^D$ is negligible compared to the first, in the $c\rightarrow \infty$ limit. So, we find :
\begin{equation} \label{Ji5tilde}
J_{i5} = \lim_{c \rightarrow \infty} J_{i5}^D = c \xi_0^\dagger \sigma_i \xi_0. 
\end{equation}
It is interesting to note that the curl of  this three-vector axial current is proportional to the extra term that appeared in the expression for the space part of the vector current (\ref{Ji1+2}). This is not strange, as the curl of an axial vector is an ordinary (polar) vector.

Following our standard algorithm, we next replace $\eta_0$ by its expression (\ref{eta-xi}). Consequently, $J^0_{5D}$ becomes :
\begin{equation}\label{J05tilde}
J_{05} = \lim_{c \rightarrow \infty}J_{05}^D = {i\hbar \over 2m} \bigg\{ {\partial \xi_0^\dagger \over \partial x_i} \sigma_i \xi_0 - \xi_0^\dagger \sigma_i {\partial \xi_0 \over \partial x_i} \bigg \}
\end{equation}
and $J_5^D$ becomes :
\begin{equation}\label{J5tilde}
J_5 = \lim_{c \rightarrow \infty} J_5^D = c\bigg \{ {\partial \xi_0^\dagger \over \partial x_i} \sigma_i \xi_0 + \xi_0^\dagger \sigma_i {\partial \xi_0 \over \partial x_i} \bigg \}.
\end{equation}
 Expressions (\ref{Ji5tilde}), (\ref{J05tilde}) and (\ref{J5tilde}) yield the structures of the chiral currents and the chiral mass term in the Pauli-Schr\"odinger theory. Let us note that $ J_{05}$ is already suppressed by a factor of $c$ compared to $ J_{i5}$ and $ J_5$. Therefore the charge density $\rho_5 = J_{05}/c $ is suppressed by a factor $c^2$ and so is 
 $\partial_0 J_{05}$. So, while computing the divergence of the axial current, we drop the term $\partial_0 J_{05}$. Now it is trivial to see that $\partial_i J_{i5}$, computed from (\ref{Ji5tilde}) precisely matches $ J_5$, as given by (\ref{J5tilde}) and the divergence equation (\ref{PCAC}) is verified using non-relativistic forms. 

It may be observed that the equality of $\partial_\mu J^{\mu D}_5$ and $J_5^D$ in the non-relativistic limit, is simply an identity and unlike the conservation of the vector current, one does not require the equations of motion to demonstrate it. The only input necessary is the relation (\ref{eta-xi}). This is different from the relativistic case, where (\ref{PCAC}) has to be obtained by exploiting the equations of motion.

A remarkable feature in the above mentioned equation is its chiral limit. If $m \rightarrow 0$, one would expect $J_5^D$ to vanish, as it is apparent from its definition (\ref{J5}). However, 
if we start with a non-vanishing rest mass, take the non-relativistic limit and then make $m \rightarrow 0$, we end up with a non-zero result as one can see from (\ref{J5tilde}). The reason is that, as $m$ becomes small, the field $\eta_0$ becomes proportionately large, as the relation (\ref{eta-xi}) indicates. The order of the limits should not be reversed, as in the massless limit, a particle tends to move at the speed of light, hence the non-relativistic limit does not make sense.
\newpage
\begin{center}
{\bf Section IV. Current Algebra }
\end{center}

\noindent {\bf IVA. Algebra of the Vector and the Axial Currents and Charges} 
\bigskip

If we quantize the Dirac theory, the fields $\psi$ and $\psi^\dagger$ obey the following equal-time anti-commutation relation :
\begin{equation}\label{Anticomm}
\{ \psi_\alpha (x), \psi^\dagger_\beta (y) \}_{x^0 = y^0} = \delta_{\alpha\beta} \delta^{(3)} (\vec x - \vec y),
\end{equation}
where $\alpha$, $\beta$ are the spinor indices.
 The bilinears, constructed out of these fields, consequently, obey the following algebra :
 \begin{equation} \label{Gamma1,2}
[\psi^\dagger(x) \Gamma_1 \psi(x), \psi^\dagger(y) \Gamma_2 \psi(y)] = \psi^\dagger(x) [\Gamma_1, \Gamma_2] \psi(x)  
\delta^{(3)} (\vec x - \vec y),
\end{equation}
where $\Gamma_1$ , $\Gamma_2$  are any two members of the Clifford algebra. By making  suitable choices for them, we can compute the commutators of our interest. For the standard Dirac vector and axial vector currents, defined before as : $J^{\mu_ D} \equiv c (\bar \psi \gamma^\mu \psi) $ and  $J^{\mu D}_5 \equiv c (\bar \psi \gamma^\mu \gamma_5 \psi) $, we find :
\begin{eqnarray}  
[J_0^D (x), J_i^D (y)] & = & [J_0^D (x), J_i^{5D}(y)]  = 0, \label{J0,Ji}\\ 
~[J_0^{D (x)}, J_{05}^D(y)] & = & 0 , \label{J0,J05}\\
~[J^{05}_D (x), J_i^D (y)] & = & [J^{05}_D (x), J_{{i5}_D} (y)] = 0. \label{J05,Ji} 
\end{eqnarray}
The non-relativistic limits of the above currents however, do not obey such trivial commutation relations. In this limit, $ J_0^D \rightarrow c (\xi_0^\dagger \xi_0)$, while 
$ J_i^D$ assumes the rather complicated expression (\ref{Ji1+2}). The limiting value of  $J_i^{5D}$, on the other hand is obtained in (\ref{Ji5tilde}) and that of  $J^0_{5D}$, is furnished in (\ref{J05tilde}). In the Pauli- Schr\"odinger Lagrangain, $i\hbar\xi_0^\dagger$ is canonically conjugate to $\xi_0$ and hence follow the anti-commutation relation :
\begin{equation}\label{Anticommxi}
\{ \xi_{0\alpha}(x), \xi_{0\beta}^\dagger(y) \}_{x^0 = y^0} = \delta_{\alpha\beta}\> \delta^{(3)} (\vec x - \vec y),
\end{equation}
when the theory is quantized. Here the spinor indices $\alpha$, $\beta$ of course, take only two values. The non-relativistic counterparts of (\ref{J0,Ji}) to (\ref{J05,Ji}) can now be obtained by straightforward calculation. The results that follow are completely based on the dictionary, connecting the relativistic and non-relativistic expressions.
The space component of the vector current $ J_i$ has two pieces, as shown in (\ref{Ji1+2}). The second piece commutes with the time component $ J_0$, while the first piece does not. The overall result for the commutator becomes :
\begin{equation} \label{J0,Ji-1} 
[J_0(x), J_i(y)] = {i\hbar c \over m} \xi_0^\dagger (y) \xi_0(y) \> \partial_i^y\delta^{(3)} (\vec x - \vec y) = {i\hbar \over m} J_0(y) \> \partial_{iy}\delta^{(3)} (\vec x - \vec y).
\end{equation}
Also, the commutator of $ J_0$ and the space component of the axial current 
$ J_{i5}$ vanishes in the non-relativistic limit, like the relativistic case :
\begin{equation} \label{J0,Ji5-1} 
[J_0(x), J_{i5}(y)] = 0.
\end{equation}
It is interesting to note that in the $ c\rightarrow  \infty $ limit, the time component of the axial current is suppressed with respect to its space component by a factor of $c$. In case of the vector current however, the time component dominates its space counterpart by the same factor! Therefore, the commutator of $J_{05}$ with $J_i $ is negligible compared to all the other commutators mentioned in (\ref{J0,Ji}), (\ref{J0,J05}) and (\ref{J05,Ji}) and may be considered to vanish.   
\begin{equation} \label{J05,Ji-1}  
[J_{05}(x), J_i(y)] = 0.
\end{equation}
The two charge operators, corresponding to the vector and the axial vector current no longer commute in the non-relativistic limit. Their commutator works out to be :
\begin{equation} \label{J0,J05-1}  
[J_0(x), J_{05}(y)] = {i\hbar \over m} J_{i5}(y) \> \partial_i^y\delta ^{(3)} (\vec x - \vec y).
\end{equation}
 The commutator of the time and the space component of the axial current reveals an interesting fact. The result reads :
\begin{eqnarray}  
[J_{05}(x), J_{i5}(y)] & = &\!\! -{i\hbar c \over m} \xi_0^\dagger (x) \xi_0(x) \> \partial^i_x \delta^{(3)} (\vec x - \vec y)  \nonumber \\
& + & \!\!\!{i\hbar c \over m}\epsilon_{ikl} [\partial_k \xi_0^\dagger (x) \sigma_l \xi_0(x) - \xi_0^\dagger (x) \sigma_l \partial_k \xi_0 (x)] \delta ^{(3)} (\vec x - \vec y) \label{J05,Ji5-1}\\
& = & \!\! -{i\hbar \over m} J_0(x) \>\partial_i^x \delta^{(3)} (\vec x - \vec y) \nonumber \\
& + & \!\!\! {i\hbar c \over m}\epsilon_{ikl} [\partial_k \xi_0^\dagger (x) \sigma_l \xi_0(x) - \xi_0^\dagger (x) \sigma_l \partial_k \xi_0 (x)] \delta^{(3)} (\vec x - \vec y). \label{J05,Ji5-2}
\end{eqnarray}
While the first term in the expression can be easily identified, the second term cannot be expressed in terms of the space or the time component of any of the currents introduced so far. 
This indicates the existence of new currents and new symmetries in the theory, 
so that the closure of the algebra is ensured.  

\bigskip
\noindent {\bf IVB. New Currents and The Extended Current algebra} 
\bigskip

The basic field in the pure Schr\"odinger theory is a single-component object, but in the Pauli-Schr\"odinger theory it is a two-component spinor. Because of this speciality, the theory permits an extra, non-abelian type symmetry transformation :
\begin{equation} \label{XiSym}
\xi_0 \rightarrow e^{-i\theta_i \sigma_i} \xi_0, 
\end{equation}
where the basic field is denoted by $\xi_0$, to maintain uniformity with the earlier discussion.
The Noether current and charge densities corresponding to this symmetry can be found following the standard prescription :
\begin{equation} \label{New J}
{\cal J}_{0k} = c\> \xi_0^\dagger \sigma_k \xi_0, \qquad  {\cal J}_{ik} = {i\hbar \over 2m}\bigg [ (\partial_i\xi_0^\dagger )\sigma_k\xi_0 - \xi_0^\dagger \sigma_k (\partial_i \xi_0) \bigg ].   
\end{equation}
The conservation of this current is guaranteed by Noether’s theorem and follows from the Pauli-Schr\"odinger equation of motion. One might be surprised to note that : $ {\cal J}_{0i}$ is identical to the $i$-th component of the (space part of) the axial current $ J_{i5}$ (\ref{Ji5tilde}) in the non-relativistic limit. On the other hand, the time component of the axial current $ J_{05}$ (\ref{J05tilde}), in the same limit, equals ${\cal J}_{ii}$, with the two indices contracted in (\ref{New J}). 
\begin{equation} \label{cal-tilde}
{\cal J}_{0k}  = J_{k5},  \qquad  {\cal J}_{ii} = J_{05}.
\end{equation}
One immediately observes that the unidentified piece in the right hand side of  (\ref{J05,Ji5-2}) is related to the new current and the commutation relation now reads :    
 \begin{equation} \label{J05,Ji5-3}
 [J_{05}(x), J_{i5}(y)]  = -{i\hbar \over m} J_0(x) \>\partial_i^x \delta^{(3)} (\vec x - \vec y) 
 + 2c \>\epsilon_{ikl} {\cal J}_{kl} (x) \>\delta^{(3)} (\vec x - \vec y),
\end{equation}
where ${\cal J}_{kl} $ is obtained from (\ref{New J}).  

The components of this new current satisfy the following commutation relations among themselves :
\begin{eqnarray}
 [{\cal J}_{0k}(x), {\cal J}_{0m}(y)] & = & 2ic \>\epsilon_{kmn} {\cal J}_{0n}(x) \>\delta^{(3)} (\vec x - \vec y), \label{NewJ0J0}\\
 ~[{\cal J}_{0k}(x), {\cal J}_{im}(y)] & = & 2ic\>\epsilon_{kmn} {\cal J}_{in}(x) \delta^{(3)} (\vec x - \vec y) \nonumber \\
& + & {i\hbar \over m} \delta_{km} J_0(y) \> \partial_i^y \>\delta^{(3)} (\vec x - \vec y).\label{NewJ0Ji}
\end{eqnarray}
To complete the story, we should compute the commutation relations of $ {\cal J}_{0k} $ and $ {\cal J}_{ik} $ with the components of the conventional vector and axial current. The results of these computations are furnished below. The time component of the new current follows the following commutation rules :
\begin{eqnarray}
  [{\cal J}_{0k}(x), J_0(y)] & = & 0, \label{NewJ0,J0}\\
  ~[{\cal J}_{0k}(x), J_{05}(y)] & = & 2ic\>\epsilon_{kmn} {\cal J}_{mn}(x) \>\delta^{(3)} (\vec x - \vec y) \nonumber \\ 
& - & {i\hbar \over m} J_0(x) \> \partial_k^x \>\delta^{(3)} (\vec x - \vec y), \label{NewJ0,J05}\\
  ~[{\cal J}_{0k}(x), J_i(y)] & = & -{i\hbar \over m} 
  {\cal J}_{0k}(x) \> \partial_i^x \>\delta^{(3)} (\vec x - \vec y) \nonumber \\
& - & {i\hbar \over m} {\cal J}_{0i}(x) \> \partial_k^x \>\delta^{(3)} (\vec x - \vec y) \nonumber \\
& + & {i\hbar \over m} \delta_{ik} \> {\cal J}_{0m} (x) \> \partial_m^x \>\delta (x - y), \label{NewJ0,Ji}\\
  ~[{\cal J}_{0k}(x), J_{i5}(y)] & = & 2ic\>\epsilon_{kil} {\cal J}_{0l}(x) \>\delta^{(3)} (\vec x - \vec y).
\end{eqnarray}
The space components  ${\cal J}_{ik}(x)$, on the other hand, exhibit the following behaviour :
 \begin{eqnarray}
[J_0(x), {\cal J}_{ik}(y)] & = & {i\hbar \over m} 
{\cal J}_{0k}(y) \> \partial_i^y \>\delta^{(3)} (\vec x - \vec y), \label{J0,NewJi}\\
~[J_{05}(x), {\cal J}_{ik}(y)] & = & 0. \label{J05,NewJi}\\
\end{eqnarray}
The last result, like (\ref{J05,Ji}), is a consequence of the non-relativistic approximation  - the commutator becomes negligible as none of the currents involved carries a factor of $c$. 

\begin{center}
{\bf Section V. Parity}
\end{center}

We are familiar with the way a Dirac spinor transforms under parity and also familiar with the fact that this transformation acts as a discrete symmetry of the Dirac Lagrangian. It is therefore interesting to find how the scenario looks in the non-relativistic limit. The parity operation on a Dirac spinor is of the form :
\begin{equation} \label{Paritypsi}
\psi(t, {\bf r}) \rightarrow \gamma^0 \psi(t, {\bf  - r}).
\end{equation}
In terms of the two-component spinors, it may be expressed as :
\begin{eqnarray}
\xi_0(t, {\bf r}) & \rightarrow& \xi_0(t, {\bf  - r}), \nonumber \\
\eta_0(t, {\bf r}) & \rightarrow& - \eta_0(t, {\bf  - r}). \label{Parityxi}
\end{eqnarray}
This is consistent with (\ref{eta-xi}), where the negative sign for the $\eta$-field is generated from the space derivative of the $\xi$-field.  Now clearly, the Lagrangian density (\ref{LDxi0}), expressed in terms of $\xi_0$ and $\eta_0$ is invariant under (\ref{Parityxi}) and so is (\ref{LPSxi0}), expressed solely in terms of $\xi_0$.

We have introduced a number of charge and current densities for the Pauli- Schr\"odinger theory. Now, a true 3-scalar should be invariant under parity, while a true 3-vector should flip sign under it. Let us check these properties for the above mentioned bi-linears. As expected, $J_0 = c \xi_0 ^\dagger \xi_0$ is invariant under (\ref{Parityxi}), while $J_i$, given by (\ref{JDixi0}), (\ref{JDixi01}), or (\ref{Ji1+2}) alters sign. The negative sign is produced by the $\eta$-field in (\ref{JDixi0}) and the space derivative in (\ref{JDixi01}) and (\ref{Ji1+2}). The behaviours remain same before and after taking the non-relativistic limit. 

In case of the axial current, $J_{05}^D$ and $J_{i5}^D$, given by (\ref{J05}) and (\ref{Ji5}), exhibit the opposite behaviour. 
\begin{equation} \label{ParityJ5}
J_{05}^D(t, {\bf r}) \rightarrow - J_{05}^D(t, {\bf r}), \quad J_{i5}^D(t, {\bf r}) \rightarrow J_{i5}^D (t, {\bf - r}).
\end{equation}
These transformation rules are maintained by (\ref{J05tilde}) and (\ref{Ji5tilde}) where the $\eta$ field is eliminated and the $c \rightarrow \infty$ limit is taken. These transformation properties demonstrate the axial nature of  $ J_{05}$ and $J_{i5}$ in the framework of the Pauli- Schr\"odinger theory. 
Furthermore, the object $ J_5^D$ defined in (\ref{J5}), or $ J_5$ in (\ref{J5tilde}) is expected to be a pseudoscalar and that nature is easily established from its behaviour under (\ref{Parityxi}). 

Finally, we turn to our newly introduced charge and current densities. ${\cal J}_{0k}$ and ${\cal J}_{ik}$ is defined by (\ref{New J}). Invoking (\ref{Parityxi}), it is trivial to check that ${\cal J}_{0k}$ is a 3-scalar and ${\cal J}_{ik}$ is a 3-vector, like $J_0$ and $J_i$ respectively.

\begin{center}
{\bf Section VI. Conclusion}
\end{center}

In this paper, we have developed a systematic algorithm for obtaining the non-relativistic limit of any function or functional of the Dirac spinors. The Dirac equation is shown to reduce to the the Pauli- Schr\"odinger equation in general, which, in absence of the magnetic field, further reduces to a simple, 
Schr\"odinger - like equation involving two-component spinors. 

The negative energy solutions of the Dirac equation are shown to satisfy a Schr\"odinger - like equation with a negative mass, which is expected. In other words, our algorithm is able to reproduce standard results for both limits, depending on which component is taken as large or small. This is reminiscent of the two types of non-relativistic (electric and magnetic) limits of the Maxwell theory.

The non-relativistic limit of the Dirac (vector) current is shown to reduce to the standard Schr\"odinger current with an additional term, which carries the signature of the spin - magnetic field coupling, even when the magnetic field is not switched on. A new, non-relativistic counterpart of the axial current of Dirac theory is obtained and the conservation of both the vector and the axial current is discussed. An interesting feature of the axial Ward identity, viz., non-vanishing of the so-called chiral mass term in the chiral limit, is revealed and justified.

The algebra of these non-relativistic currents are worked out and found to have a far more rich structure than that in the relativistic Dirac theory. In particular, the algebra does not close, hinting at the existence of new currents and new symmetries. These new currents and the underlying symmetries are unearthed and the full algebra, including these new currents are furnished. 

We have also discussed about the form the parity operator assumes in the non-relativistic limit. The symmetry of the non-relativistic theory under this transformation is demonstrated. The behaviour of the above mentioned currents under parity is discussed and it is established that the newly introduced axial current, indeed, has an axial nature.    

The present work opens up a number of areas for future investigation. Now that we have an axial current along with the vector one, one can study chiral anomaly within a {\it non-relativistic} framework, if one quantizes the Pauli- Schr\"odinger theory. One can also work out the current algebra afresh in this quantum version of the theory and look for the so-called commutator anomalies. 

The non-abelian type symmetries, mentioned in section IVB, of the Pauli- Schr\"odinger theory can be gauged and the phenomenological consequences of that model can be investigated. The Pauli- Schr\"odinger theory has an intrinsic non-abelian nature, but it can be further enriched by coupling the spinor field with a non-abelian gauge field.

\end{document}